\documentclass[epj]{webofc}
\usepackage[utf8]{inputenc}
\usepackage[varg]{txfonts}   
\usepackage{booktabs}
\usepackage{xcolor}
\definecolor{darkred}{rgb}{0.4,0.0,0.0}
\definecolor{darkgreen}{rgb}{0.0,0.4,0.0}
\definecolor{darkblue}{rgb}{0.0,0.0,0.4}
\usepackage[bookmarks,linktocpage,colorlinks,
    linkcolor = darkred,
    urlcolor  = darkblue,
    citecolor = darkgreen]{hyperref}
\usepackage{subfigure}
\wocname{EPJ Web of Conferences}
\woctitle{Lattice2017}

\newcommand*{\bea}{\begin{eqnarray}}
\newcommand*{\eea}{\end{eqnarray}}
\newcommand*{\be}{\begin{equation}}
\newcommand*{\ee}{\end{equation}}

\newcommand{\bma}{\begin{pmatrix}}
\newcommand{\ema}{\end{pmatrix}}

\begin{document}

\selectlanguage{english}

\title{
Gluon and ghost correlation functions of 2-color QCD at finite density
}



\author{%
\firstname{Ouraman} \lastname{Hajizadeh}\inst{1}\fnsep\thanks{Speaker, \email{ouraman.hajizadeh@uni-graz.at}} \and
\firstname{Tamer} \lastname{Boz}\inst{2} \and
\firstname{Axel}  \lastname{Maas}\inst{1} \and 
\firstname{Jon-Ivar}  \lastname{Skullerud}\inst{2}
}
\institute{%
Institute of Physics, NAWI Graz, University of Graz, Austria
\and
Department of Theoretical Physics, Maynooth University, Ireland
}

\abstract{2-color QCD, i.\ e.\ QCD with the gauge group SU(2), is the simplest non-Abelian gauge theory without sign problem at finite quark density. Therefore its study on the lattice is a benchmark for other non-perturbative approaches at finite density. To provide such benchmarks we determine the minimal-Landau-gauge 2-point and 3-gluon correlation functions of the gauge sector and the running gauge coupling at finite density. We observe no significant effects, except for some low-momentum screening of the gluons at and above the supposed high-density phase transition.}

\maketitle

\section{Introduction}

Understanding the full phase diagram of QCD from first principles
studies has been a challenge for decades. The most interesting
phenomena occur in the strongly interacting regime, where perturbative
methods fail. Therefore non-perturbative approaches like functional
methods and lattice gauge theory are applied to study QCD in the
different thermodynamic regimes. Unfortunately, lattice QCD suffers
from the sign problem at finite density. Therefore, it does not so far
seem possible to reproduce at finite density the interplay of
lattice and functional methods that has proven fruitful in the finite temperature case~\cite{Maas:2011se}. One way to circumvent this problem is the study of QCD-like theories \cite{Kogut:2000ek} at finite density on the lattice, which do not suffer from the sign problem. Of course, these theories need to share as many properties with real QCD as possible to be really constraining. The simplest such theory is QC$_2$D, i.\ e.\ QCD with the gauge group SU(2) rather than SU(3), but otherwise left unchanged. Therefore, QC$_2$D has already been extensively studied, especially using lattice methods \cite{Hands:2006ve,Hands:2010gd,Cotter:2012mb,Boz:2013rca,Braguta:2015owi,Scior:2015vra,Braguta:2016cpw,Wellegehausen:2017gba,Holicki:2017psk}. The focus of the present contribution is on the properties of the gauge sector at finite density, especially the (minimal) Landau-gauge propagators and 3-point vertices and the running gauge coupling. These are of prime importance as inputs and/or benchmarks for functional calculations \cite{Maas:2011se}. 

Our results indicate no strong change with respect to the vacuum. Only
around the supposed deconfinement phase transition and at very high
density do we see the onset of a moderate screening of the gluons. If this finding would be generic, this would considerably simplify studies of finite densities using functional methods, as then the gauge sector could be left essentially as in the vacuum, as has been done, e.\ g., already in \cite{Nickel:2006kc,Nickel:2006vf,Marhauser:2006hy,Nickel:2008ef,Contant:2017gtz}, supporting the findings in these studies. Conversely, this implies that the high-density phase remains strongly coupled in the whole range of densities studied here.

To set the stage, we will briefly rehearse the pertinent features of the zero-temperature phase diagram of QC$_2$D in section \ref{s:pd}, and sketch our methods in section \ref{s:tech}. We then present our results for the correlation functions in section \ref{s:gluon}-\ref{s:g3v}.

\section{Phase diagram}\label{s:pd}

The phase diagram of QC$_2$D has been studied extensively in \cite{Hands:2006ve,Hands:2010gd,Cotter:2012mb,Boz:2013rca,Braguta:2015owi,Braguta:2016cpw,Wellegehausen:2017gba,Holicki:2017psk}. As in the following the configurations used in \cite{Hands:2006ve,Hands:2010gd,Cotter:2012mb,Boz:2013rca} will be employed, we will describe here this particular case with two degenerate flavors of Wilson fermions. We concentrate here essentially only on the low(zero)-temperature and finite density regime. There, it was found that after the silver blaze point a phase arose with thermodynamic properties close to that of a free gas, up to about four times the silver blaze chemical potential. At that point another transition occurred where the thermodynamic properties are quite distinct from a free gas, and in which also quantities like a large Polyakov loop signal a quite different physics. The latter could be interpreted as a would-be deconfinement.

In addition, here also the main difference to QCD arises. In QC$_2$D it is possible to form a gauge-invariant bosonic diquark, though not a fermionic baryon. Since the diquark is bosonic, it can condense, and does so above the silver-blaze point, forming a superfluid medium. Thus, at first sight there seems to be first a weakly interacting phase followed by a non-trivial, but 'deconfined' phase. However in the termodynamically trivial phase, the Wilson potential is not compatible with a weakly interacting phase \cite{Boz:2013rca}.

\section{Setup and methods}\label{s:tech}

As noted, we use in the following the ensembles from \cite{Boz:2013rca, Cotter:2012mb} with $\beta=1.9$ and $\kappa= 0.168$, corresponding to a lattice spacing $a=0.178(6)$ fm, and a pion mass $m_\pi=717(25)$ MeV on a $16^3\times 24$ lattice. In this case, a non-zero diquark source is introduced to force the diquark condensate as the expected physical phase. Other lattice parameters will be studied elsewhere \cite{BHMS:unpublished}. These configurations were gauge-fixed to minimal Landau gauge using the methods described in \cite{Cucchieri:2006tf}. The propagators, the running coupling, and the 3-point vertices were then determined using the methods described in \cite{Cucchieri:2007ta,Fister:2014bpa}. The gluon propagator at finite density was already determined in \cite{Boz:2013rca,Hands:2006ve}, and the independent determination here is in agreement. Further technical details will also be available elsewhere \cite{BHMS:unpublished}, but are also reviewed in \cite{Maas:2011se}.

\section{Gluon propagator}\label{s:gluon}

\begin{figure}[thb]
\includegraphics[width=0.5\textwidth]{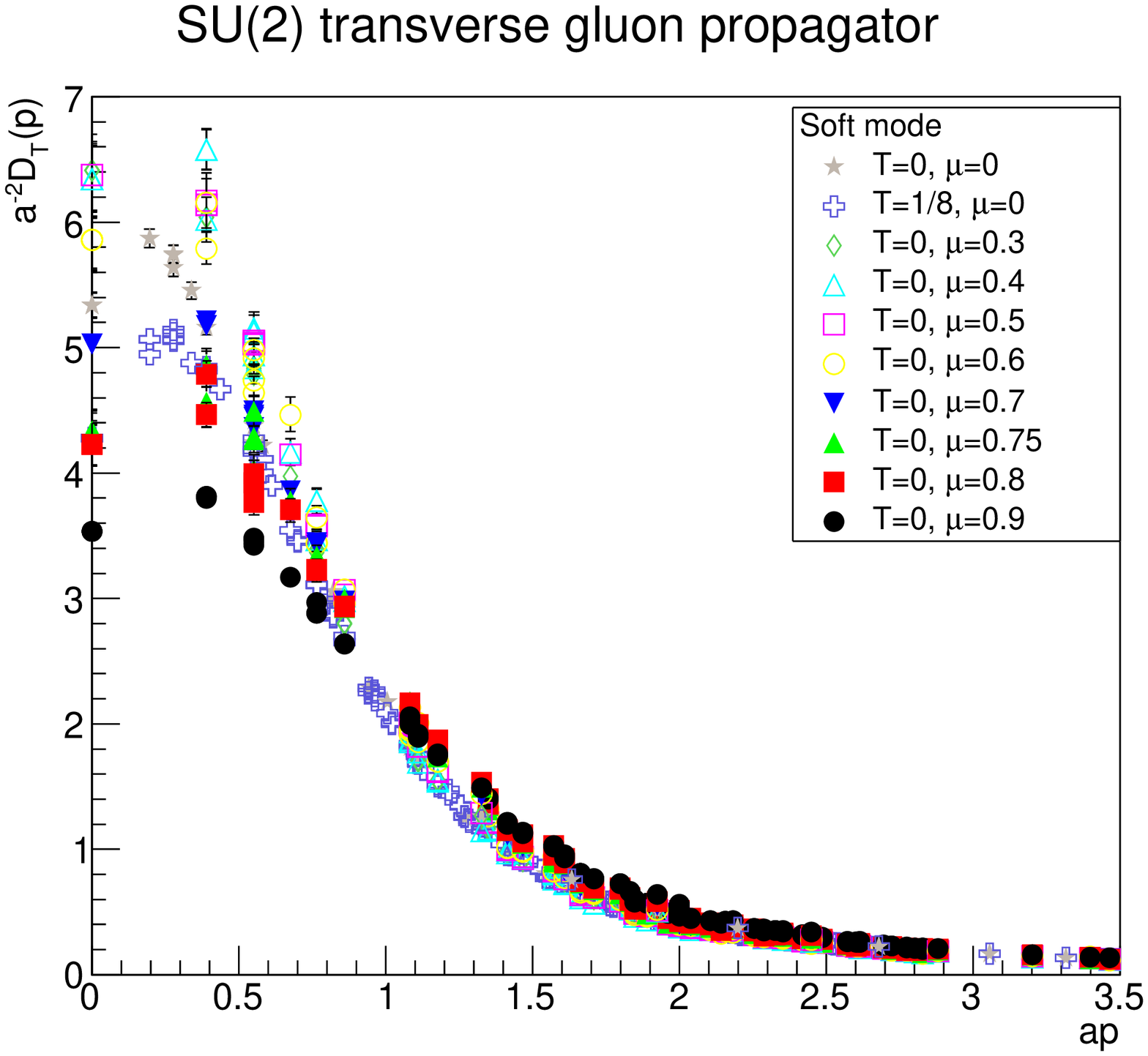}\includegraphics[width=0.5\textwidth]{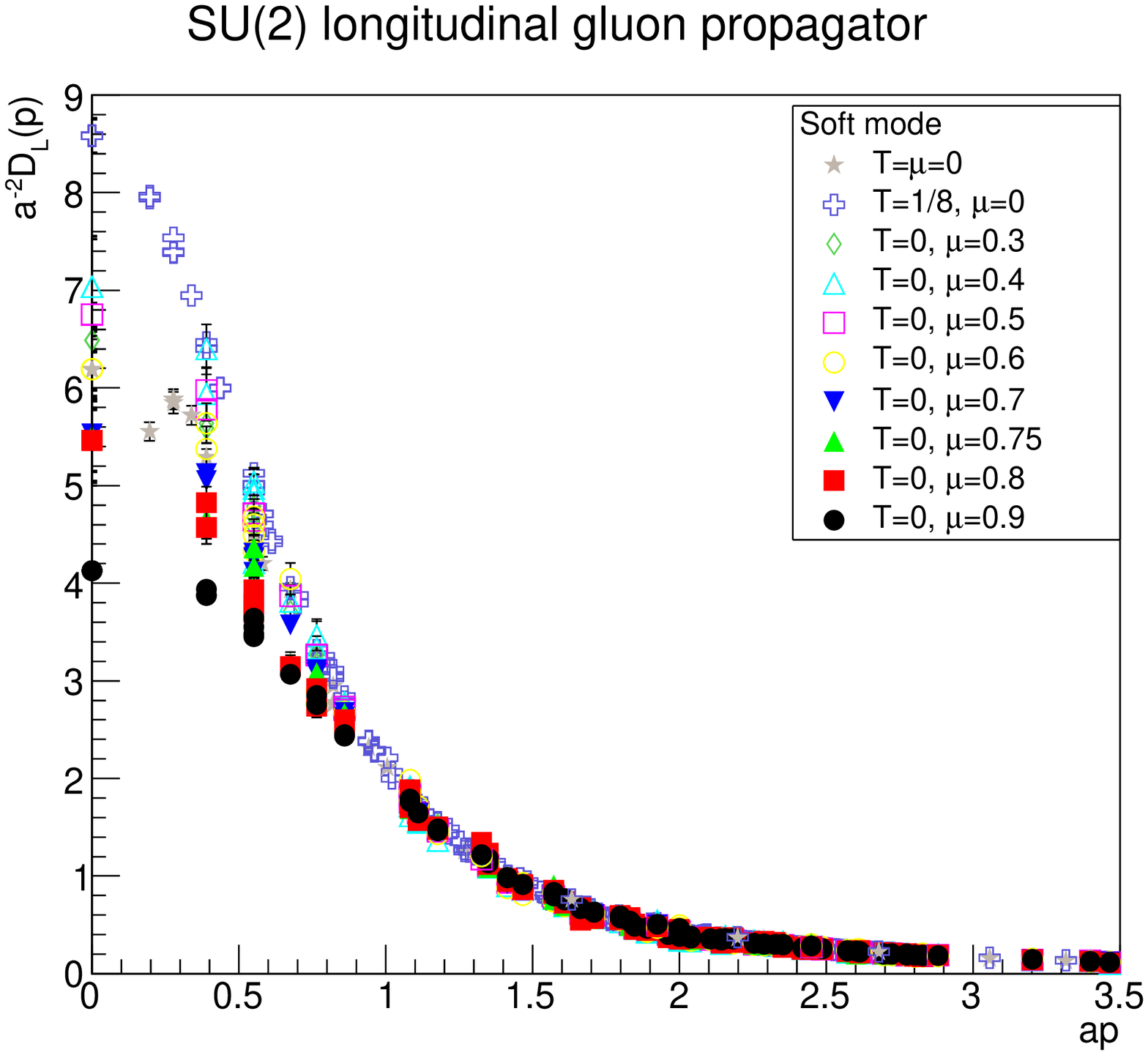}
\includegraphics[width=0.5\textwidth]{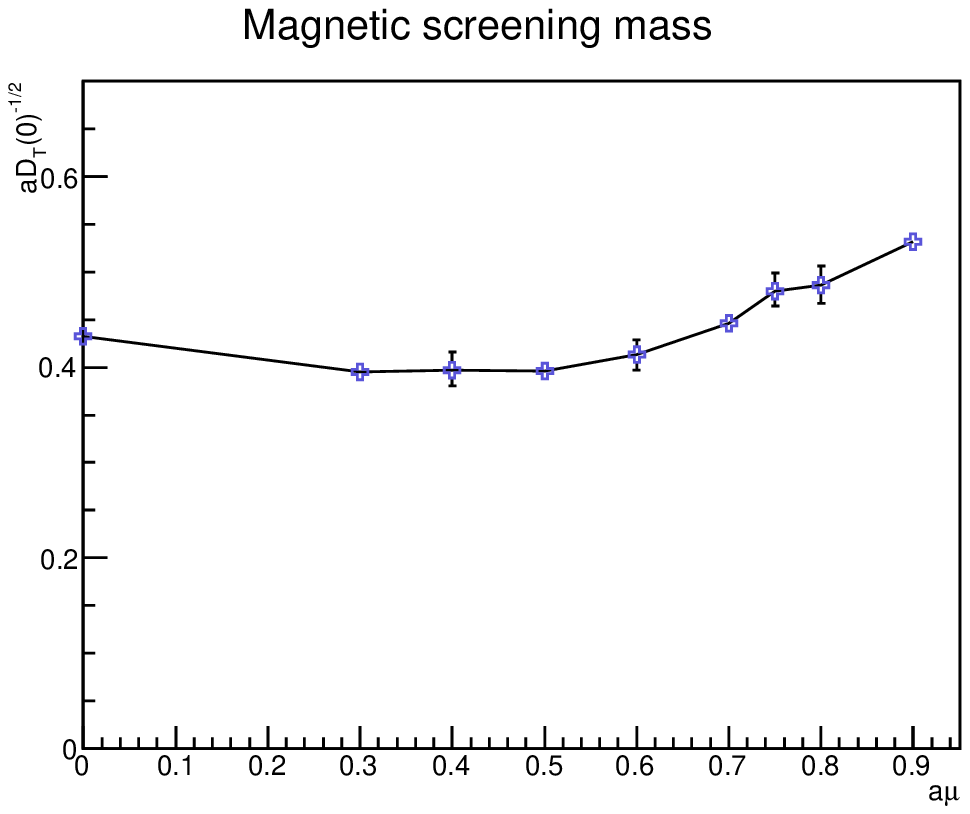}\includegraphics[width=0.5\textwidth]{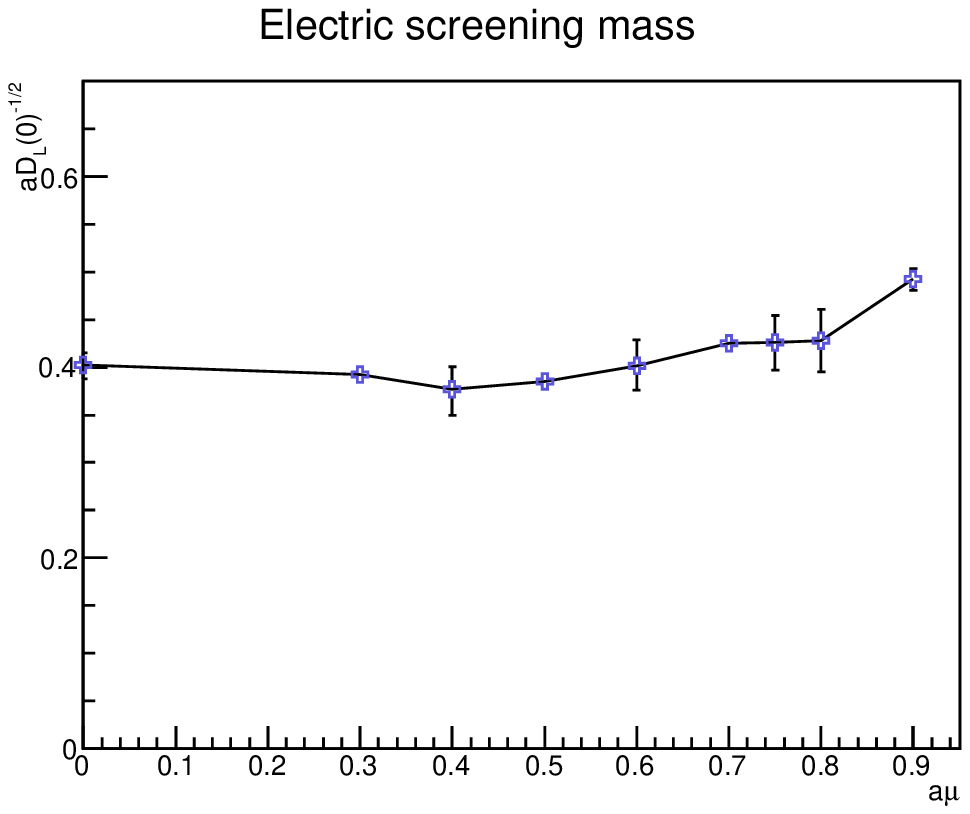}
\caption{\label{fig:gpmu}
 The soft mode of the gluon propagator perpendicular (top-left panel, 'magnetic') and parallel (top-right panel, 'electric') to the heat-bath \cite{Maas:2011se} as a function of momentum for multiple densities. The lower panel shows the respective screening masses, as discussed in the text. For comparison results in the vacuum and at finite temperature are also displayed. If not visible, (statistical) error bars are smaller than the symbol sizes. Note that, because only a single lattice spacing was used, no renormalization has been performed.} 
\end{figure}

The results for the soft mode of the gluon propagator are shown in
figure \ref{fig:gpmu}. In agreement with previous studies
\cite{Boz:2013rca,Hands:2006ve}, we do not see any indication of a
density dependence below the deconfinement phase transition. Starting
from the phase transition we do see an infrared suppression, which
increases with density. This is also seen in the corresponding
screening masses, defined as $1/\sqrt{D(0)}$, where $D(p)$ is the
propagator. This screening mass is essentially constant above the
silver-blaze point, and only starts to increase slowly with the
chemical potential above the phase transition. Thus, the contribution from the gluons themselves to thermodynamics would be unaltered below the transition, and only slightly changed above the transition.

We also do not see a marked difference between the electric and magnetic propagator.

Both observations are in stark contrast to the situation above the finite-temperature phase transition \cite{Aouane:2012bk,Maas:2011ez,Cucchieri:2014nya,Silva:2017feh}. The inertness below the transition, is however, similar. But in the finite-temperature case the inertness is not accompanied by an almost free thermodynamics.

There is also a slight modification of the magnetic propagator seen at intermediate momenta, which is not observed in the electric one. However, the effect is small, and it is not beyond doubt whether this could be a lattice artifact.

\section{Ghost propagator} \label{s:ghost}

\begin{figure}
\begin{center}
\includegraphics[width=0.5\textwidth]{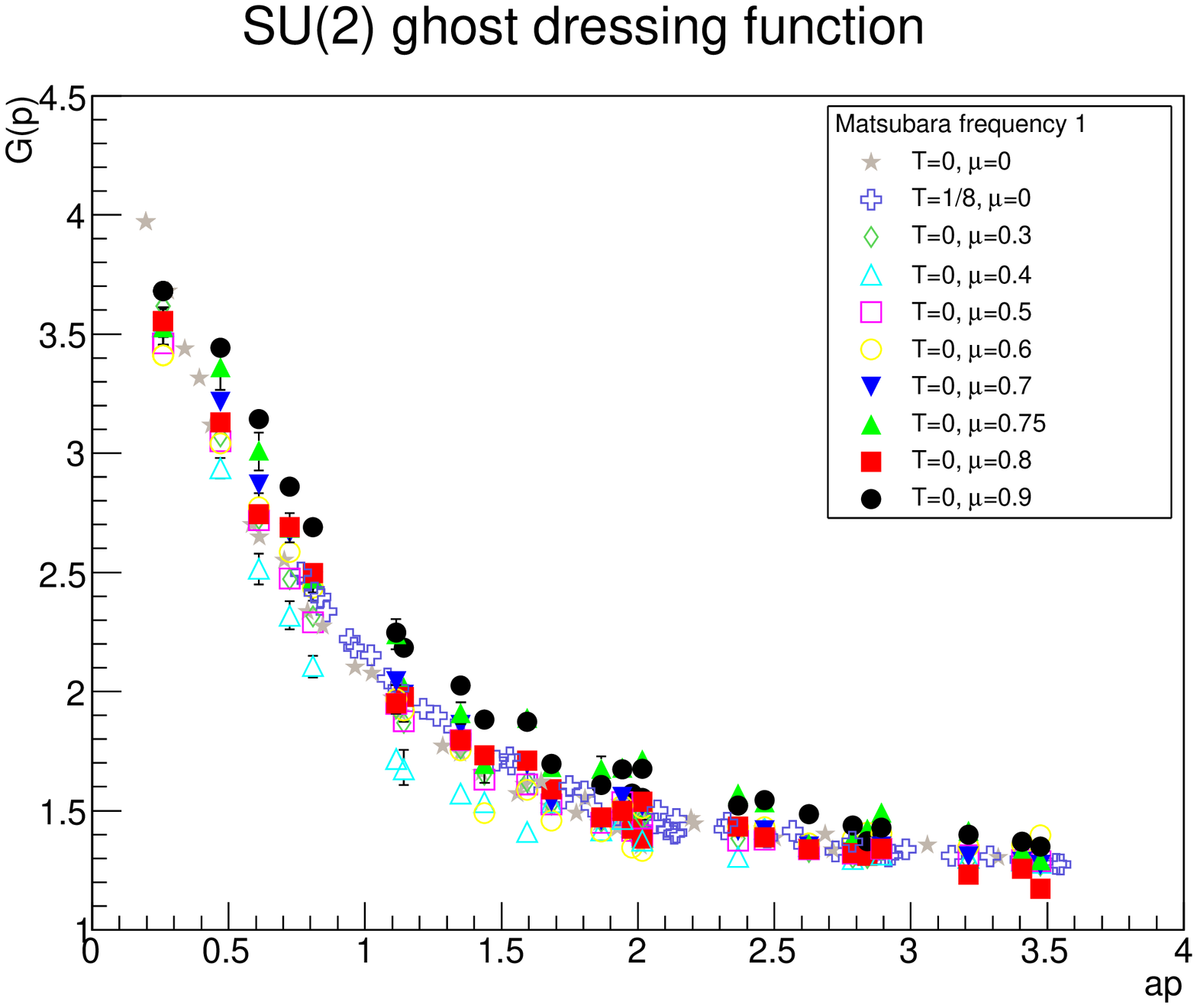}\includegraphics[width=0.5\textwidth]{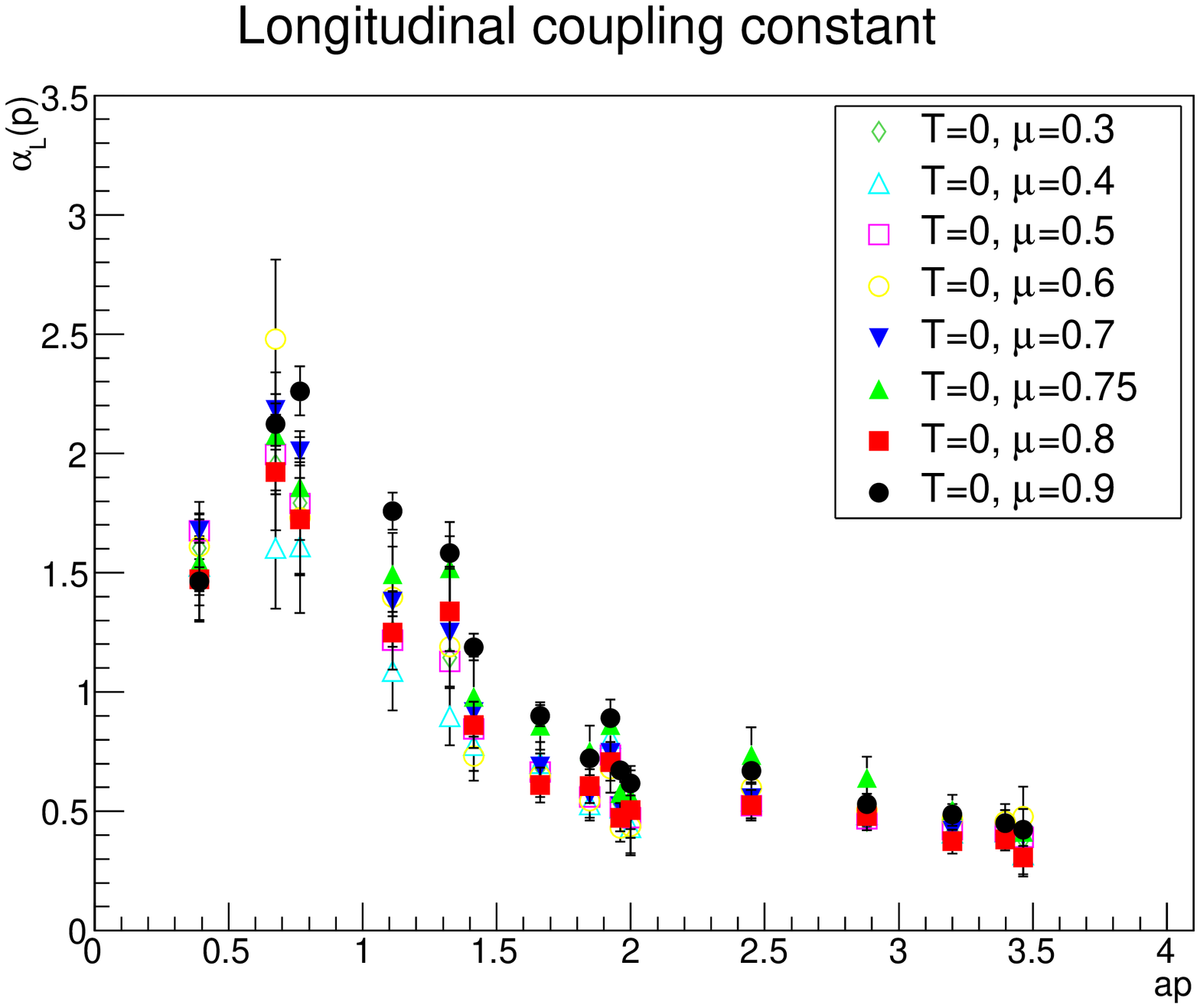}
\end{center}
\caption{\label{fig:ghd}The soft mode of the ghost dressing function, i.\ e.\ ghost propagator multiplied with momentum squared, as a function of momentum for multiple densities (left panel) and the electric running coupling in the miniMOM scheme as a function of momentum for multiple densities (right panel). For comparison for the ghost dressing function also results in the vacuum and at finite temperature are displayed. If not visible, (statistical) error bars are smaller than the symbol sizes. Note that, because only a single lattice spacing was used, no renormalization has been performed.} 
\end{figure}

The soft mode of the ghost propagator, shown in figure \ref{fig:ghd} (left), does not show any systematic dependence on the chemical potential. Some effects are seen, but since the data do
not show any visible trend, they are likely some type of not yet fully understood lattice artifact. This behavior is in agreement with what is seen at finite temperature \cite{Aouane:2012bk,Maas:2011ez}, where the propagator is also inert to temperature effects.

Using the gluon propagator and the ghost propagator together, it is
possible to determine the running gauge coupling in the miniMOM scheme
\cite{vonSmekal:2009ae}. However, because of the splitting of electric
and magnetic degrees of freedom, there are now two different running
couplings, characterizing the interactions of magnetic and electric
degrees of freedom. However, because both do not behave significantly
differently, the corresponding couplings are also similar. The
electric running coupling is shown in figure \ref{fig:ghd} (right). As
a consequence of the inertness of the propagators, it is essentially
density-independent. This is a very remarkable result. As this running
coupling also governs the interaction between quarks and gluons this
implies that the microscopic interaction remains essentially
unchanged, even though in the one phase almost trivial thermodynamics
prevail and in the other phase a would-be deconfinement is
observed. However, this fits very well with the observation that the
Wilson potential and Polyakov loop in the thermodynamically trivial phase still show a confining behavior \cite{Hands:2006ve,Hands:2010gd,Cotter:2012mb,Boz:2013rca}. Still, this is a very remarkable observation.

\section{Vertices}\label{s:g3v}
 
At finite temperature a surprisingly strong and qualitative response of the magnetic, soft three-gluon vertex on the phase transition was seen in exploratory lattice simulations \cite{Fister:2014bpa}. We therefore study it at finite density.

\begin{figure}
\begin{center}
\includegraphics[width=\textwidth]{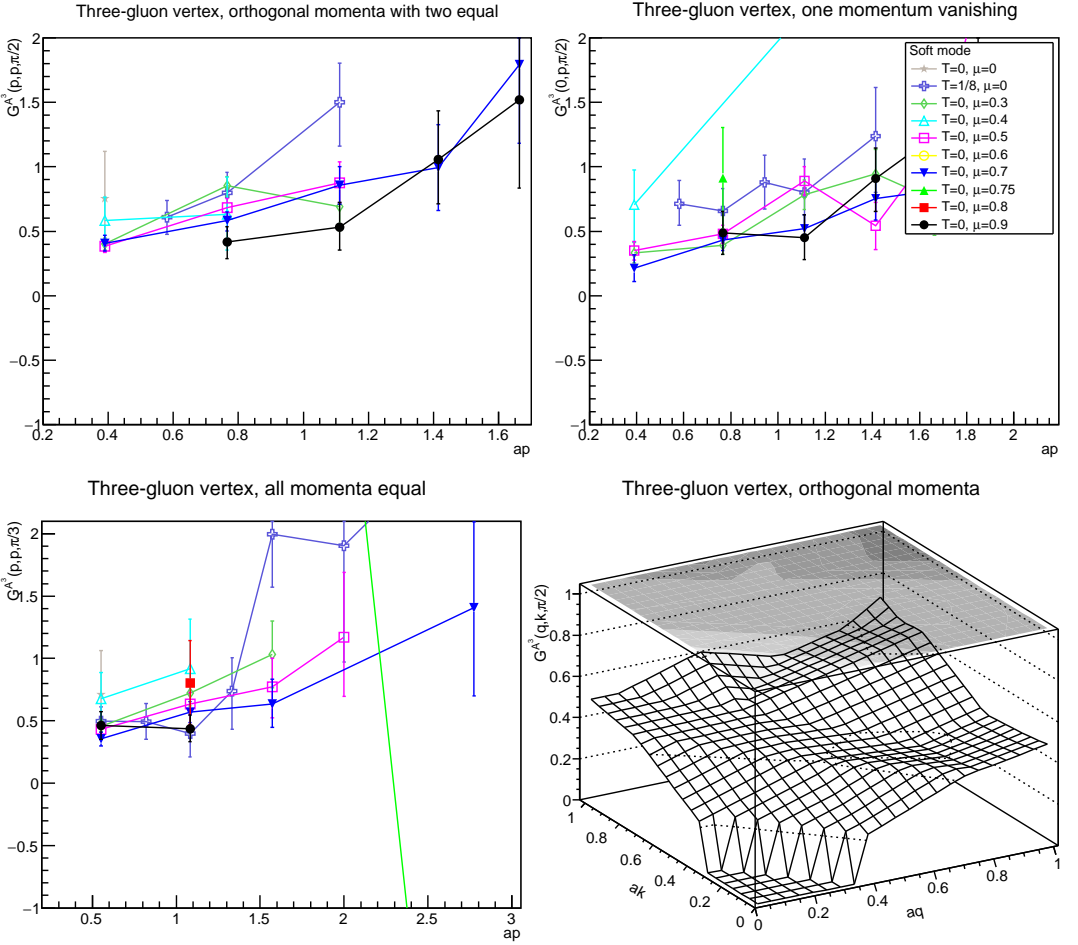}
\end{center}
\caption{\label{fig:3gv-mu}The tree-level dressing function of the
  soft magnetic three-gluon vertex as a function of momentum for
  multiple densities. For comparison also results in the vacuum and at
  finite temperature are displayed. The momentum configurations
  displayed are two of the gluon momenta being orthogonal momenta with
  equal size in the top-left panel, one momentum vanishing in the
  top-right panel, and all momenta equal, i.\ e.\ the symmetric
  momentum configuration, in the lower-left panel. The
  three-dimensional plot in the lower-right panel gives an
  interpolation of the data for two momenta being orthogonal at the
  chemical potential close to the "deconfinement" transition, at $\mu a=0.7$. We only display points for which our statistics were sufficient to reach 50\% or less statistical error. Lines are drawn to guide the eye.}
\end{figure}
 
The result of studying three gluon vertex is limited due to large statistical noise. The data points with reasonable statistical errors are shown in figure \ref{fig:3gv-mu}. Though noisy, it is seen that within statistical accuracy no distinction is observed. In particular, the distinct infrared suppression observed in lattice calculations at zero temperature and density \cite{Cucchieri:2008qm,Duarte:2016ieu,Sternbeck:2017ntv} remains.

We also studied the soft magnetic tree-level dressing function of the ghost-gluon vertex which shows, as at finite temperature \cite{Fister:2014bpa}, no marked dependence on the density. These results will be presented in more detail elsewhere \cite{BHMS:unpublished}.

\section{Conclusions and outlook}

Summarizing, we studied the gauge sector's propagators and three-point vertices as a function of density in QC$_2$D. In the low-density regime we did not find any influence of the density on the correlation functions. Above the supposed "deconfinement" phase transition \cite{Hands:2006ve,Hands:2010gd,Cotter:2012mb,Boz:2013rca} we do observe some moderate additional screening for the gluon propagator. However, no effect is seen in either the ghost sector nor for the magnetic three-gluon vertex. Also, the influence on the magnetic and electric sector is, within errors, the same. This is in marked contrast to the finite-temperature case, where both sectors react differently. Also, the lack of influence on the magnetic three-gluon vertex is noteworthy, as this is again in contrast to the finite-temperature case. Finally, the running coupling does essentially not change as a function of the density. Thus, the system remains microscopically strongly coupled for all densities.

These findings only add to the enigma of the finite-density physics of this theory \cite{Hands:2006ve,Hands:2010gd,Cotter:2012mb,Boz:2013rca}: The low-density phase shows almost trivial, non-interacting thermodynamic properties, while both the Wilson potential and the findings here indicate strong microscopic interactions. While a similar phenomenon is seen in the magnetic sector at very high temperatures, here it is seen in both magnetic and electric sectors. Thus it cannot be explained in the same way. In the high-density phase the theory remains strongly coupled and thermodynamically non-trivial, but here quantities like the Polyakov loop show the same behavior as at very high temperatures.

Thus, a full picture of the theory remains elusive. Complementary investigations using functional methods may help understand it better. For these, our results can act as a benchmark or allow to justify the approximation to keep the gauge sector unaltered compared to the vacuum, except for additional screening due to quarks \cite{Nickel:2006kc,Nickel:2006vf,Marhauser:2006hy,Nickel:2008ef,Contant:2017gtz}.

Finally, lattice artifacts can be expected to be relevant, given the experiences at finite temperature \cite{Maas:2011se,Aouane:2012bk,Maas:2011ez,Cucchieri:2014nya,Silva:2017feh}. Thus, a more detailed study is necessary, which will be presented elsewhere \cite{BHMS:unpublished}.

\section*{Acknowledgements}

O.\ H.\ was supported by the FWF DK W1203-N16. We acknowledge the networking support by the COST action CA15213 ``Theory of hot matter and relativistic heavy-ion collisions''.

\bibliographystyle{woc}
\bibliography{bib}

\end{document}